\documentclass[preprint,apjl,dvipdfmx]{aastex}

\usepackage{natbib}
\shorttitle{Gravitation Instability in Disk of Porous Dust}
\shortauthors{Michikoshi \& Kokubo}

\begin{document}

\title{
  Planetesimal Formation by Gravitational Instability of a Porous-Dust Disk
}
\author{
  Shugo Michikoshi\altaffilmark{1}, and Eiichiro Kokubo\altaffilmark{2}
}
\altaffiltext{1}{
    Center for Computational Sciences, University of Tsukuba, Tsukuba, Ibaraki 305-8577, Japan
  }
\altaffiltext{2}{
  Division of Theoretical Astronomy, National Astronomical Observatory of
  Japan, Osawa, Mitaka, Tokyo 181-8588, Japan
}

\email{
  michikos@ccs.tsukuba.ac.jp, and kokubo@th.nao.ac.jp
}

\begin{abstract}
Recently it is proposed that porous icy dust aggregates are formed by pairwise accretion of dust aggregates beyond the snowline.
We calculate the equilibrium random velocity of porous dust aggregates taking into account mutual gravitational scattering, collisions, gas drag, and turbulent stirring and scattering. 
We find that the disk of porous dust aggregates becomes gravitationally unstable as they evolve through gravitational compression in the minimum-mass solar nebula model for a reasonable range of turbulence strength, which leads to rapid formation of planetesimals.
\end{abstract}

\section{Introduction}
In the standard scenario of planet formation, planetesimals are the building
blocks of planets \citep[e.g.,][]{Safronov1969, Hayashi1985}.
In a protoplanetary disk small dust grains grow to km-sized objects called planetesimals.   
From planetesimals, protoplanets or planetary embryos form through a process of the runaway and oligarchic growth \citep[e.g.,][]{Kokubo1998, Kokubo2012}.
However, the formation mechanism of planetesimals is one of today's most important unsolved problems.

In the classical model of planetesimal formation, the gravitational
 instability (GI) plays a key role.
As dust particles grow large and decouple from gas, they settle onto
 the disk midplane and form a dust layer.  
When the density of the dust layer exceeds the Roche density, the GI 
 occurs \citep{Safronov1969, Goldreich1973, Hayashi1985}.  
The gravitationally unstable dust disk fragments into gravitationally
 bound objects \citep{Michikoshi2007, Michikoshi2009, Michikoshi2010}. 
They finally become planetesimals as they shrink.

The above model assumes no turbulence in the gas disk.
However, the turbulence is likely to be driven by the magneto-rotational
 instability \citep[e.g.,][]{Sano2000} and the shear instability
 \citep[e.g.,][]{Sekiya2000, Michikoshi2006}. 
Under the turbulence dust particles are stirred up and cannot settle onto
the midplane \citep[e.g.,][]{Weidenschilling1993}. 
In the standard minimum mass solar nebular (MMSN) model, the GI does not
 occur \citep{Sekiya1998}. 
One of the possible mechanisms for overcoming this difficulty is the
 streaming instability, which leads to the formation of the gravitationally bound objects \citep{Youdin2005, Johansen2007}.
 
Another formation model is the pairwise coagulation of dust particles. 
The recent studies on the dust growth showed that the icy dust aggregates
 formed by coagulation are not compact but significantly porous
 \citep{Dominik1997,Blum2000,Wada2007,Wada2008,Wada2009,Suyama2008,Suyama2012,Okuzumi2012}.  
The internal density of dust aggregates is much smaller than the
 material density, which is $\sim 10^{-5}\,\mathrm{g}\,\mathrm{cm}^{-3}$. 
Some compression mechanism is necessary to form compact planetesimals with
 $\sim 1\,\mathrm{g}\,\mathrm{cm}^{-3}$. 
\cite{Kataoka2013} found that the dust aggregates can be compressed by
 the ram pressure when the dust aggregate mass $m_\mathrm{d} \lesssim 10^{11}\, \mathrm{g}$.  
The dust aggregates with $m_\mathrm{d} \gtrsim 10^{11}\, \mathrm{g}$ are
 compressed by the self-gravity and finally the density reaches
 $\sim 0.1\,\mathrm{g}\,\mathrm{cm}^{-3}$.
 
In this paper, we revisit the final stage of the dust aggregate
 evolution by the gravitational compression.
We investigate the dynamics of the porous dust aggregates and demonstrate
 that the GI takes place as a result of the dust evolution.
In Section \ref{sec:method} we describe the calculation method.
We present the results in Section \ref{sec:res}.
Section \ref{sec:sum} is devoted to a summary and discussion.

\section{Model and Method \label{sec:method}}

\subsection{Disk Model}

We adopt the surface densities of gas and dust,
 $\Sigma_\mathrm{g} =
 1700 f_\mathrm{g} (a/\mathrm{AU})^{-3/2} \, \mathrm{g}\, \mathrm{cm}^{-2}$ and
 $\Sigma_\mathrm{d} = f_\mathrm{d} \Sigma_\mathrm{g}$, where $a$ is the
 distance from the central star, $f_\mathrm{g}$ is the ratio to the MMSN
 model and $f_\mathrm{d}=0.018$ is the dust-to-gas mass ratio beyond the
 snowline \citep{Hayashi1981, Hayashi1985}.
We adopt the temperature profile
 $T = T_1 (a/\mathrm{AU})^{-3/7} \,\mathrm{K}$,
 where $T_1=120$ \citep{Chiang2010}.
The isothermal sound velocity is
 $c_\mathrm{s}=\sqrt{k_\mathrm{B}T/m_\mathrm{g}}$, where $k_\mathrm{B}$
 is the Boltzmann constant and
 $m_\mathrm{g} = 3.9 \times 10^{-24} \, \mathrm{g}$ is the mean molecular mass.
The gas density at the disk midplane is
 $\rho_\mathrm{g} = \Sigma_\mathrm{g}/(\sqrt{2 \pi}  c_\mathrm{s}/\Omega)$, where $\Omega = \sqrt{G M_*/a^3}$ is the Keplerian frequency and $M_*$ is the
 central star mass. 
 We adopt $M_* = M_\odot$.
The mean free path of gas molecules is
 $l = m_\mathrm{g} / \sigma_\mathrm{g} \rho_\mathrm{g}$, where
 $\sigma_\mathrm{g} = 2 \times 10^{-15}\, \mathrm{cm}^2$ is the
 collisional cross-section of gas molecules. 
The nondimensional radial pressure gradient is given as
$\eta = -(1/2)[c_\mathrm{s}/(a \Omega)]^2 \partial \log (\rho_\mathrm{g}
 c_\mathrm{s}^2) / \partial \log a$. 

We consider the spherical porous dust aggregate with mass $m_\mathrm{d}$ and radius $r_\mathrm{d}$, consisting of monomers with radius $r_0$ and density $\rho_0$.
As a first step, we assume that all the dust aggregates have the same mass \citep[e.g.,][]{Kataoka2013}. This assumption is justified if the
size distribution has a steep single peak \citep[e.g.,][]{Okuzumi2011, Okuzumi2012}.
We define the mean internal density $\rho_\mathrm{int}  = m_\mathrm{d}/(4\pi r_\mathrm{d}^3/3)$.
The geometric cross-section of the dust aggregate is given as $\pi r_\mathrm{d}^2$.

\subsection{Random Velocity of Dust Aggregates}

We calculate the equilibrium random velocity $v$ of dust aggregates considering gravitational scattering, collisions, and interaction with gas. 
For simplicity we assume the isotropic velocity distribution, that is, $v_x \simeq v_y \simeq v_z \simeq v/\sqrt{3}$, where $v_x$, $v_y$, and $v_z$ are the $x$, $y$, and $z$ components of the random velocity, respectively.

\subsubsection{Gravitational Scattering}

The random velocity increases by mutual gravitational scattering.
The timescale of gravitational scattering is well described by Chandrasekhar's relaxation time \citep{Ida1990}.
The heating rate due to gravitational scattering is 
\begin{equation}
  \left(\frac{\mathrm{d}v^2}{\mathrm{d}t}\right)_\mathrm{grav} = n_\mathrm{d} \pi \left(\frac{2Gm_\mathrm{d}}{v_\mathrm{rel}^2}\right)^2 v_\mathrm{rel} v^2 \log \Lambda , 
 \label{eq:velo}
\end{equation}
where $v_\mathrm{rel} \simeq \sqrt{2} v$ is the typical relative velocity between dust aggregates, $n_\mathrm{d} \simeq (\Sigma_\mathrm{d}/m_\mathrm{d})/(\sqrt{2 \pi}v_z/\Omega)$
is the number density of dust aggregates, and $\Lambda = v_\mathrm{rel}^2 (v_\mathrm{z}/\Omega + r_\mathrm{H})/(2Gm_\mathrm{d})$ where $r_\mathrm{H}=(2m_\mathrm{d}/3M_*)^{1/3}a $ is the Hill radius \citep{Stewart2000}. 

\subsubsection{Collision}

We assume that all collisions lead to accretion.
Under this assumption the collisional damping rate is given as
\begin{equation}
 \left(\frac{\mathrm{d}v^2}{\mathrm{d}t}\right)_\mathrm{col} = - C_\mathrm{col} n_\mathrm{d} \pi (2r_\mathrm{d})^2
 \left(1 +\frac{v_\mathrm{esc}^2}{v_\mathrm{rel}^2} \right) v_\mathrm{rel} v^2,
 \label{eq:col}
\end{equation}
 where $v_\mathrm{esc} = \sqrt{2 G m_\mathrm{d}/r_\mathrm{d}}$ is the surface escape velocity and $C_\mathrm{col}$ is the ratio of change of the kinetic energy on the collision.
 We consider that the orbit of the merged dust aggregate is given by that of the center of the mass and adopt $C_\mathrm{col}=1/2$ \citep{Inaba2001}. 
\subsubsection{Gas Effects}

We consider the three interactions between turbulent gas and dust
 aggregates, namely, drag from the mean gas flow, turbulent stirring
 due to gas drag, and gravitational scattering by the turbulent density
 fluctuations.  

The drag from the mean gas flow reduces $v$ on the stopping timescale
 $t_\mathrm{s}$ as 
\begin{equation}
 \left(\frac{\mathrm{d} v^2}{\mathrm{d}t}\right)_\mathrm{gas,drag} = -\frac{2}{t_\mathrm{s}} v^2,
 \label{eq:dragdamp}
\end{equation}
 where $t_\mathrm{s}$ is 
\begin{equation}
 t_\mathrm{s} = \frac{2 m_\mathrm{d}}{\pi C_\mathrm{D} r_\mathrm{d}^2 \rho_\mathrm{g} u}, 
 \label{eq:stop}
\end{equation}
 where $C_\mathrm{D}$ is the dimensionless drag coefficient and
 $u$ is the relative velocity between dust and gas.
We adopt the typical relative velocity
 $u \simeq \sqrt{v^2 + \eta^2 v_\mathrm{K}^2 }$, where
 $v_\mathrm{K} = a \Omega$ is the Keplerian velocity. 

The gas drag law changes with $r_\mathrm{d}$ \citep[e.g.,][]{Adachi1976}.
If $r_\mathrm{d} \gtrsim l$, we use the Stokes or Newton drag.
For the low Reynolds number case ($\mathrm{Re} \ll 10^3$), the drag
 coefficient is approximated by $C_\mathrm{D} \simeq 24/\mathrm{Re}$
 (Stokes drag), where $\mathrm{Re}=2 r_\mathrm{d} u/\nu$. 
The viscosity $\nu$ is given by $\nu = v_\mathrm{th} l/2$ where
 $v_\mathrm{th}=\sqrt{8/\pi} c_\mathrm{s}$ is the thermal velocity. 
 For the high Reynolds number case ($10^3 < \mathrm{Re} < 2 \times 10^5$), the drag coefficient is almost constant $C_\mathrm{D} \simeq 0.4 \mbox{--} 0.5$
 (Newton drag). 
If $r_\mathrm{d} \lesssim l$, we use the Epstein drag.
Thus, we adopt the drag coefficient formula as \citep{Brown2003}
\begin{equation}
C_\mathrm{D} = \left\{ \begin{array}{ll}
	\displaystyle {\frac{8 v_\mathrm{th}}{3 u}}  & (r_\mathrm{d} < 9l/4) \\
	\displaystyle {\frac{0.407}{1+8710/ \mathrm{Re}}+ \frac{24}{\mathrm{Re}}(1+0.150 \mathrm{Re}^{0.681}) } & (r_\mathrm{d} > 9l/4) 
\end{array} \right.. 
\end{equation}

In the turbulent gas, turbulence stirs dust aggregates by gas drag.
In this case $v$ reaches the equilibrium value \citep{Youdin2007}
\begin{equation}
 v^2 = \frac{v_\mathrm{t}^2t_\mathrm{e}}{ t_\mathrm{e} + t_\mathrm{s}},
 \label{eq:eqrandturb}
\end{equation}
 where $t_\mathrm{e}$ is the eddy turnover time,
 $v_\mathrm{t} = \sqrt{\alpha} c_\mathrm{s}$ is the magnitude of the turbulent velocity, and
 $\alpha$ is the dimensionless turbulence strength \citep{Cuzzi2001}. 
Thus the heating rate due to turbulent stirring is
\begin{equation}
 \left(\frac{\mathrm{d} v^2}{\mathrm{d}t}\right)_\mathrm{turb,stir} = \frac{2 \tau_\mathrm{e} v_\mathrm{t}^2 \Omega}{S(\tau_\mathrm{e} + S)},
 \label{eq:dragturb}
\end{equation}
 where $\tau_\mathrm{e} = t_\mathrm{e} \Omega$ and $S = \Omega t_\mathrm{s}$ is the Stokes number. 
We adopt $\tau_\mathrm{e} = 1$ \citep{Youdin2011, Michikoshi2012}.

The gas density fluctuates because of the turbulence.
The dust aggregates are gravitationally scattered by the density fluctuations.
\cite{Okuzumi2013} considered the magneto-rotational instability
 turbulence and derived the fitting formula of the heating rate
\begin{equation}
 \left(\frac{\mathrm{d} v^2}{\mathrm{d}t}\right)_\mathrm{turb,grav} = C_\mathrm{turb} \alpha \left(\frac{\Sigma_\mathrm{g} a^2}{M_*} \right)^2
 \Omega^3 a^2, 
\end{equation}
where $C_\mathrm{turb}$ is the dimensionless factor that depends on the disk structure.
We assume that the dead zone thickness is comparable to that of gas and adopt $C_\mathrm{turb} = 3.1 \times 10^{-2}$ \citep{Okuzumi2013}.

\subsubsection{Equilibrium Random Velocity}

The evolution of $v$ is described as
\begin{equation}
 \frac{\mathrm{d}v^2}{\mathrm{d}t} =
 \left(\frac{\mathrm{d} v^2}{\mathrm{d}t}\right)_\mathrm{grav}
 + \left(\frac{\mathrm{d} v^2}{\mathrm{d}t}\right)_\mathrm{col}
 + \left(\frac{\mathrm{d} v^2}{\mathrm{d}t}\right)_\mathrm{gas,drag} 
 + \left(\frac{\mathrm{d} v^2}{\mathrm{d}t}\right)_\mathrm{turb,stir} 
 + \left(\frac{\mathrm{d} v^2}{\mathrm{d}t}\right)_\mathrm{turb,grav}.
 \label{eq:random}
\end{equation}
We can calculate the equilibrium random velocity of dust aggregates by setting $\mathrm{d}v^2/\mathrm{d}t = 0$. 

\subsection{Gravitational Instability Condition}

To investigate the dynamical stability of the disk of dust aggregates, we
 use Toomre's $Q$ 
\begin{equation}
 Q = \frac{v_x \Omega}{3.36 G \Sigma_\mathrm{d}},
 \label{eq:qvalue}
\end{equation}
 with the equilibrium random velocity \citep{Toomre1964}.
For the axisymmetric mode, the instability condition is $Q < 1$ \citep{Toomre1964}. 
   However, for $1 \lesssim Q \lesssim 2$, the non-axisymmetric mode or self-gravity wakes can grow on the dynamical timescale 
   \citep[e.g.,][]{Toomre1981, Salo1995, Michikoshi2007, Michikoshi2016}. 
   \cite{Michikoshi2007, Michikoshi2009, Michikoshi2010} showed that the wakes fragment to form planetesimals. Therefore we adopt the
   condition $Q < Q_\mathrm{crit} = 2$.

Note that in the regime of $Q \lesssim Q_\mathrm{crit}$ near the evolution track of
self-gravitational compression in our model (see Section 3.1), we obtain
$S \gg 1$, and thus the secular GI reduces to the dynamical GI \citep{Youdin2011, Takahashi2014}.

\section{Results \label{sec:res}}

\subsection{Evolution of Dust Aggregates}

\begin{figure}
  \begin{center}
	\plotone{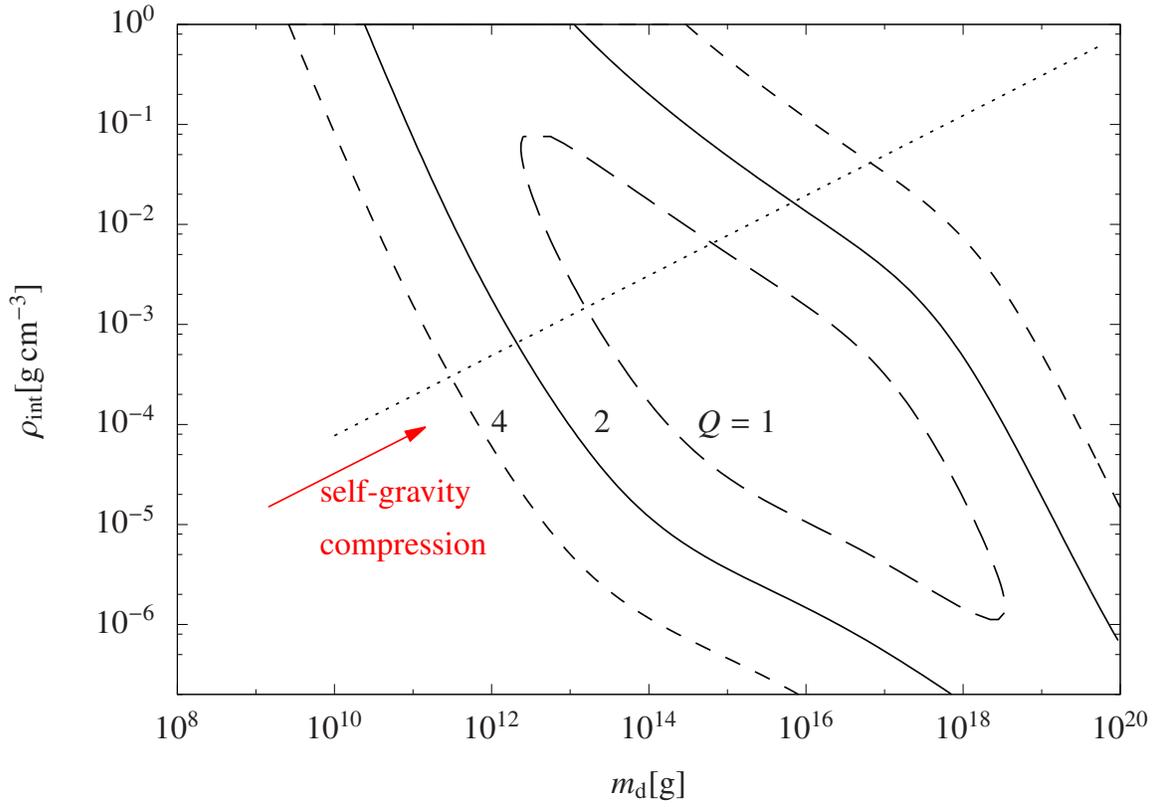}
  \end{center}
  \caption{GI region on the $m_\mathrm{d}$--$\rho_\mathrm{int}$ plane at $5 \, \mathrm{AU}$ for the fiducial model. The dashed, solid, and short-dashed curves show contours for $Q=1, 2,$ and $4$, respectively.
	The dotted line represents the evolution track of dust aggregates by the self-gravitational compression \citep{Kataoka2013}.
  }
      \label{fig:qvalplot_ok}
\end{figure}

We calculate $v$ and then $Q$ for a disk of porous dust aggregates with
 $m_\mathrm{d}$ and $\rho_\mathrm{int}$.
Figure \ref{fig:qvalplot_ok} shows $Q$ on the
 $m_\mathrm{d}$-$\rho_\mathrm{int}$ plane for the fiducial model at
 $5\,\mathrm{AU}$, where $f_\mathrm{g} = 1$ and $\alpha = 10^{-3}$. 
We find a wide GI region with $Q < Q\mathrm{crit}$. 

We consider the evolution of dust aggregates with
 $m_\mathrm{d} \gtrsim 10^{11} \,\mathrm{g}$, where they are compressed
 by their self-gravity \citep{Kataoka2013}.
\cite{Kataoka2013} investigated the evolution in this regime
 considering the compressive strength
 $P_\mathrm{comp}=E_\mathrm{roll} \rho_\mathrm{int}^3 /r_0^3 \rho_0^3$
 and the self-gravitational pressure
 $P_\mathrm{grav}=Gm_\mathrm{d}^2/\pi r_\mathrm{d}^4$, where
 $E_\mathrm{roll}$ is the rolling energy. 
We draw the evolution track of dust aggregates in Figure
 \ref{fig:qvalplot_ok}, assuming 
 $E_\mathrm{roll}=4.74\times10^{-9}\, \mathrm{erg}$, $\rho_0=1.0 \,
 \mathrm{g}\, \mathrm{cm}^{-3}$, and $r_0=0.1 \, \mu \mathrm{m}$. 
The evolution track crosses the GI region. 
In other words, the porous-dust disk becomes gravitationally unstable to fragment to form
 planetesimals. 

Figure \ref{fig:source} shows the main heating and cooling mechanisms of
 the dust disk in the fiducial model.
On the evolution track for $m_\mathrm{d} \lesssim 10^{14}\, \mathrm{g}$,
 the main heating mechanism is turbulent stirring. 
Along the evolution, $r_\mathrm{d}$ increases with $m_\mathrm{d}$ as
 $r_\mathrm{d} \propto m_\mathrm{d}^{1/5}$.
For the Stokes drag, $S$ ($\propto m_\mathrm{d}/r_\mathrm{d}$)
 increases with $m_\mathrm{d}$.  
As $S$ increases, dust aggregates decouple from turbulent gas,
 which reduces their random velocity.
Therefore, $Q$ decreases with increasing $m_\mathrm{d}$ and finally
 becomes less than $Q_\mathrm{crit}$.  
 
\begin{figure}
	 \begin{tabular}{c}
		\begin{minipage}{0.5\hsize}
			\begin{center}
			  (a)
			  \includegraphics[width=\textwidth]{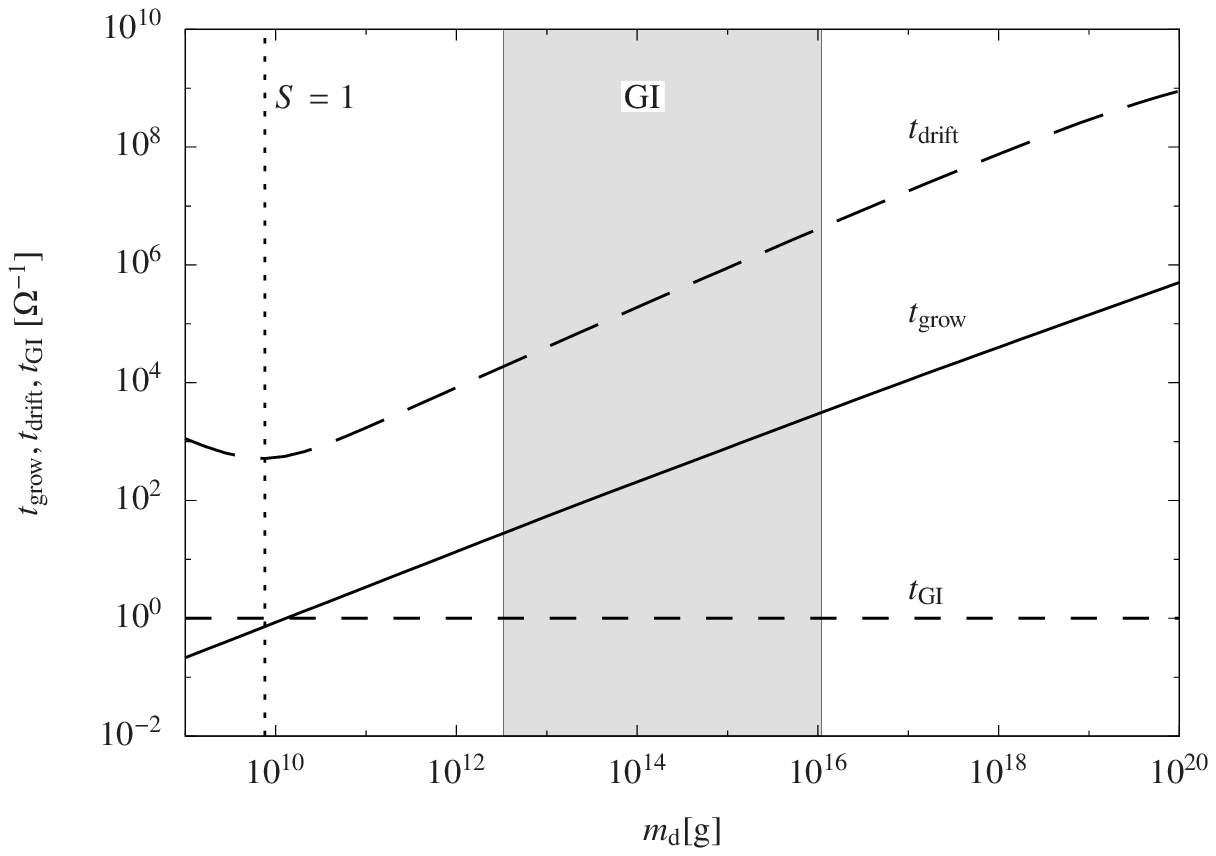}
			\end{center}
		\end{minipage}
		\begin{minipage}{0.5\hsize}
			\begin{center}
			  (b)
				\includegraphics[width=\textwidth]{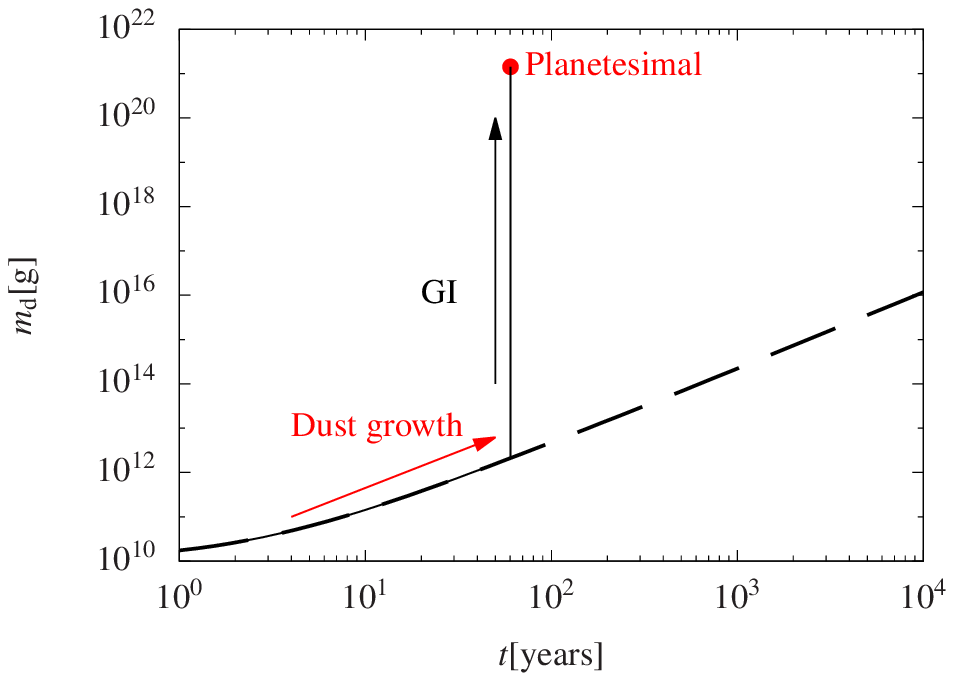}
			\end{center}
		\end{minipage}
	  \end{tabular}
  \caption{(a) Timescales of the dust growth (solid), radial drift (dashed), and GI (short dashed) in the fiducial model at $5\, \mathrm{AU}$.
	The dotted line and the shaded region correspond to $S=1$ and the GI region ($Q<Q_\mathrm{crit}$), respectively.
	(b) Time evolution of the dust mass.
The dashed curve shows the evolution only by the collisional growth along the evolution track and the solid curve shows the evolution including the GI.
The time $t$ means the time elapsed since $m=10^{10} \, \mathrm{g}$.
  }
  \label{fig:timescale}
\end{figure}

Figure \ref{fig:timescale}a shows the various timescales.
We calculate the timescales assuming the evolution track of the self-gravitational compression.
The growth time for $S>1$ is
 $t_\mathrm{grow} = m_\mathrm{d}/(\rho_\mathrm{d} \pi r_\mathrm{d}^2 v)$ where
 $\rho_\mathrm{d} = m_\mathrm{d} n_\mathrm{d}$ and we neglect
 gravitational focusing. 
The radial drift time is given as
  $t_\mathrm{drift}=a/(2 S \eta v_\mathrm{K}/(1+S^2))$
  \citep{Adachi1976, Weidenschilling1977}. 
The GI timescale is about $t_\mathrm{GI}\sim \Omega^{-1}$.
The GI is much faster than the other processes.
Thus the GI takes place once the GI condition is satisfied. 
The mass evolution of dust aggregates is shown in Figure \ref{fig:timescale}b.  
The GI immediately forms planetesimals from dust aggregates.
Note that the growth time here is under the assumption of perfect
accretion for the sake of simplicity. The realistic growth of such
huge porous dust aggregates is poorly understood.

\subsection{Disk Condition for Gravitational Instability}

\begin{figure}
	 \begin{tabular}{c}
		\begin{minipage}{0.5\hsize}
			\begin{center}
			  (a)
			  \includegraphics[width=\textwidth]{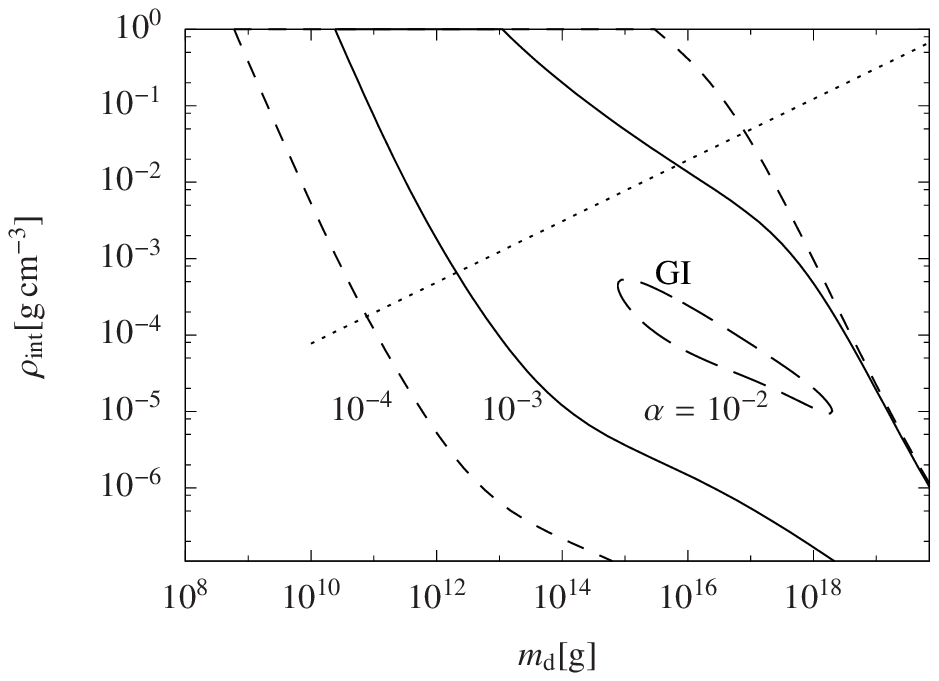}
			\end{center}
		\end{minipage}
		\begin{minipage}{0.5\hsize}
			\begin{center}
			  (b)
			  \includegraphics[width=\textwidth]{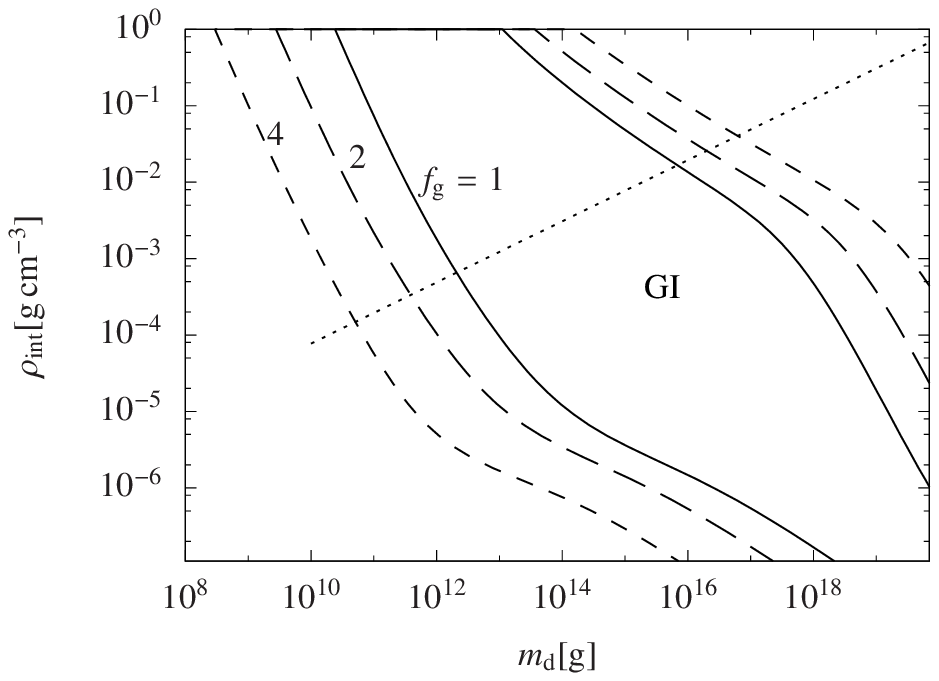}
			\end{center}
		\end{minipage}
	  \end{tabular}
	  \caption{Dependence of the GI regions on (a) $\alpha$ and (b) $f_\mathrm{g}$. (a) $\alpha = 10^{-2}$ (dashed), $10^{-3}$ (solid), and $10^{-4}$ (short-dashed) with $f_\mathrm{g}=1$.
	(b) $f_\mathrm{g} = 1$ (solid), $2$ (dashed), and $4$ (short-dashed) with $\alpha=10^{-3}$.
	The dotted line is the evolution track.
	}
      \label{fig:dependence}
\end{figure}

We investigate the disk condition for the GI.
Figure \ref{fig:dependence}a represents the dependence
 of the GI region on $\alpha$ on the $m_\mathrm{d}$-$\rho_\mathrm{int}$
 plane.  
The GI region with $\alpha=10^{-4}$ is larger than that in the fiducial
 model.  
Because the turbulence is the main source to increase the random
 velocity, $v$ is smaller for smaller $\alpha$.
Therefore, the GI region expands.
On the other hand, for $\alpha=10^{-2}$, the GI region shrinks. 
The strong turbulence suppresses the GI.
Figure \ref{fig:dependence}b represents the dependence on $f_\mathrm{g}$. 
The GI region is wider for larger $f_\mathrm{g}$.
For the massive disk, the GI more easily takes place.

We examine if the GI occurs along the dust evolution for disks with
 various $f_\mathrm{g}$ and $\alpha$.
The results are summarized in Figure \ref{fig:gicond}a.
As expected, the GI is more prone to occur for larger $f_\mathrm{g}$ and
 smaller $\alpha$.
In the MMSN model ($f_\mathrm{g} = 1$), $\alpha$ should be less than $7 \times 10^{-3}$ for the GI.  
If $f_\mathrm{g} \gtrsim 1.3$, even though the strong turbulence case ($\alpha=10^{-2}$), the GI is possible.
 
Next, we examine the dependence on $a$ with the MMSN model
 ($f_\mathrm{g} = 1$). 
Figure \ref{fig:gicond}b shows the results. 
We find that the occurrence of the GI barely depends on $a$.
The GI region exists for any $a$ if $\alpha \lesssim 1 \times 10^{-2}$. 
The upper bound of $\alpha$ for the GI, where the GI region exists and the evolution track touches it, slightly decreases with increasing $a$.
However, its dependence is weak.
For $\alpha \lesssim 5\times 10^{-3}$, the GI occurs for $a < 20\,\mathrm{AU}$. 

In all the cases where the dust evolution leads to the GI in Figure \ref{fig:gicond}, 
$t_\mathrm{GI} < t_\mathrm{grow}$ and $t_\mathrm{GI} < t_\mathrm{drift}$ are satisfied
 if the dust aggregates evolve by the self-gravitational compression.
Thus the GI is inevitable on the course of dust evolution for the above disk conditions.

\subsection{Critical Turbulence Strength}

\begin{figure}
  \begin{center}
	\plotone{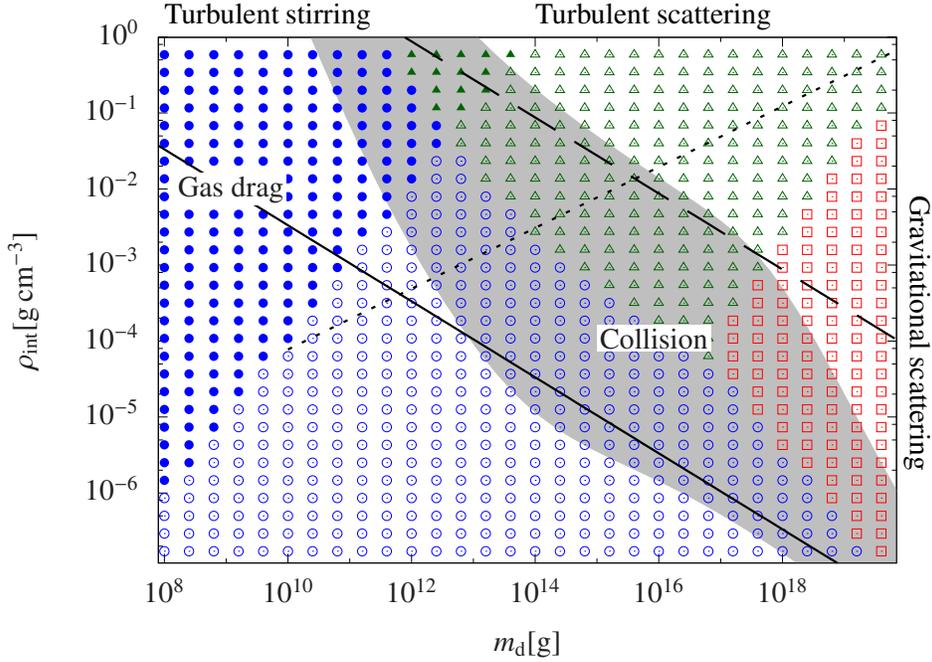}
  \end{center}
  \caption{Dominant heating and cooling mechanisms for the fiducial model at $5\,\mathrm{AU}$. 
  The filled and open symbols represent gas drag and collisional damping for the dominant cooling process, respectively. 
  The squares, circles, and triangles represent gravitational scattering, turbulent stirring, and turbulent scattering, respectively.
  The shaded region denotes the GI region where $Q<Q_\mathrm{crit}$.
  The solid and dashed lines represent the approximated instability condition described by Equations (\ref{eq:mlow}) and (\ref{eq:mhigh}), respectively.
  The dotted line represents the evolution track.
	}
      \label{fig:source}
\end{figure}

We derive the condition for the existence of the GI region as a function
 of disk parameters.
In Figure \ref{fig:source}, on the lower left boundary of the GI region,
 the main heating mechanism is turbulent stirring and the main
 cooling mechanism is collisional damping. 
Thus, we calculate $v$ from
 $(\mathrm{d} v^2/\mathrm{d}t)_\mathrm{turb,stir} + (\mathrm{d} v^2/\mathrm{d}t)_\mathrm{col}=0$ assuming
 $t_\mathrm{s} \gg t_\mathrm{e}$ and $u \simeq \eta v_\mathrm{K}$ and
 neglecting gravitational focusing.  
We obtain the condition for $Q<Q_\mathrm{crit}$ as
\begin{equation}
 m_\mathrm{d} \gtrsim m_\mathrm{low}=9.52\times 10^{-8} \frac{\alpha^3 C_\mathrm{D}^6 \eta^6 v_\mathrm{K}^6 c_\mathrm{s}^6 \rho_\mathrm{g}^6 \tau_\mathrm{e}^3 }{C_\mathrm{col}^3 Q_\mathrm{crit}^6 \rho_\mathrm{int}^2 \Sigma_\mathrm{d}^9 G^6}.
 \label{eq:mlow}
\end{equation}

On the upper right boundary of the GI region, the main heating source is
 turbulent scattering and the main cooling source is collisional damping. 
 Thus, we calculate $v$ from
 $(\mathrm{d} v^2/\mathrm{d}t)_\mathrm{turb,grav} + (\mathrm{d} v^2/\mathrm{d}t)_\mathrm{col}=0$.
The condition for the GI in this regime is 
\begin{equation}
 m_\mathrm{d} \lesssim m_\mathrm{high}=4.10 \times 10^6 \frac{C_\mathrm{col}^3 Q_\mathrm{crit}^6 \Sigma_\mathrm{d}^9}{\alpha^3 C_\mathrm{turb}^3 \rho_\mathrm{int}^2 \Sigma_\mathrm{g}^6}
 \label{eq:mhigh}
\end{equation}

\begin{figure}
	 \begin{tabular}{c}
		\begin{minipage}{0.5\hsize}
			\begin{center}
			  (a)
			  \includegraphics[width=\textwidth]{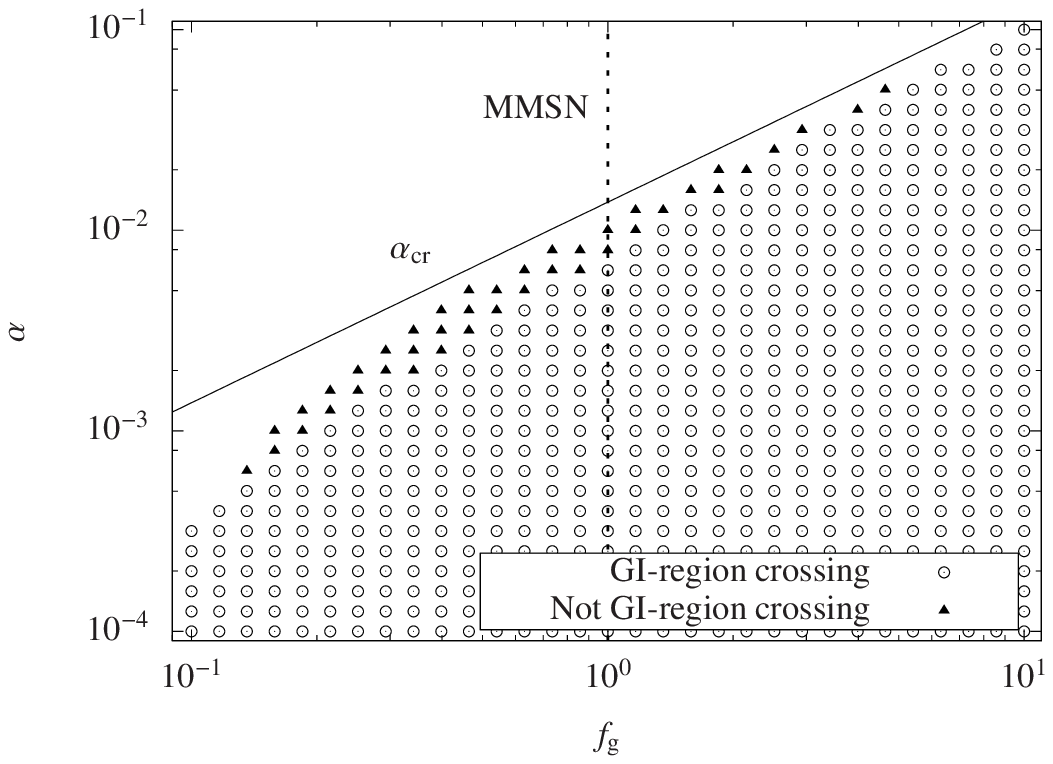}
			\end{center}
		\end{minipage}
		\begin{minipage}{0.5\hsize}
			\begin{center}
			  (b)
			  \includegraphics[width=\textwidth]{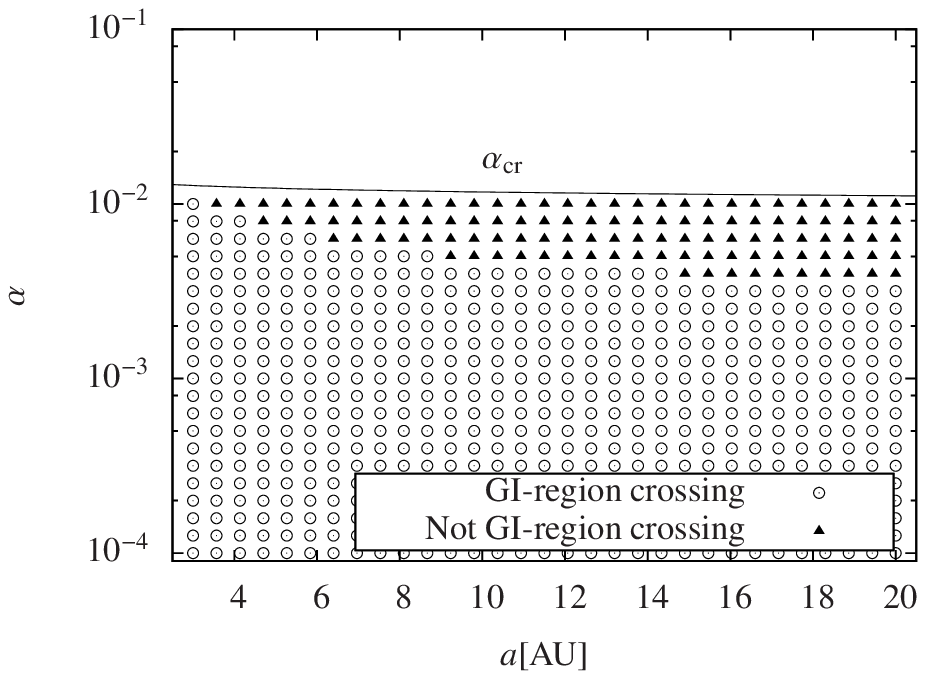}
			\end{center}
		\end{minipage}
	  \end{tabular}
  \caption{Parameter regime for the GI on the $f_\mathrm{g}$--$\alpha$ plane at $a=5\, \mathrm{AU}$ (a) and on the $a$--$\alpha$ plane with the MMSN model $f_\mathrm{g}=1$ (b).
	The points show the cases where the GI region exists in the area $10^8 < m_\mathrm{d} < 10^{20} \mathrm{g}$ and $10^{-6} \, \mathrm{g}\, \mathrm{cm}^{-3} < \rho_\mathrm{int} < 1 \, \mathrm{g}\, \mathrm{cm}^{-3}$ on the $m_\mathrm{d}$--$\rho_\mathrm{int}$ plane.
	The filled triangle represents the case where the evolution track does not cross the GI region.
	The circle shows the case where the evolution track crosses the GI region.
	The solid line represents $\alpha_\mathrm{cr}$ described by Equation (\ref{eq:cond}).  
	The dotted line corresponds to the MMSN model.
  }
      \label{fig:gicond}
\end{figure}

As shown in Figure \ref{fig:source}, these two conditions agree well
 with the numerical results.
Thus, the necessary condition for the existence of the GI region is
 $m_\mathrm{low} < m_\mathrm{high}$. 
From this, we derive the critical $\alpha$ as
\begin{equation}
 \alpha < \alpha_\mathrm{cr} = 4.70\times10^2 \frac{ C_\mathrm{col} Q_\mathrm{crit}^2 a^2 \Sigma_\mathrm{d}^3}{\sqrt{C_\mathrm{turb}\tau_\mathrm{e}} C_\mathrm{D} \eta M_* \Sigma_\mathrm{g}^2 }.
\end{equation}
Using the disk model, we rewrite $\alpha_\mathrm{cr}$ as
\begin{equation}
 \alpha_\mathrm{cr} = 1.38 \times 10^{-2} \tau_\mathrm{e}^{-1/2} f_\mathrm{g} \left(\frac{f_\mathrm{d}}{0.018}\right)^3 \left(\frac{T_1}{120}\right)^{-1} \left(\frac{C_\mathrm{turb}}{3.1 \times 10^{-2}} \right) ^{-1/2}  \left(\frac{Q_\mathrm{crit}}{2} \right) ^{2}  \left(\frac{a}{5\, \mathrm{AU}} \right) ^{-1/14},
\label{eq:cond}
\end{equation}
 where we adopt $C_\mathrm{D} = 0.5$.
The dependence of $\alpha_\mathrm{cr}$ on $a$ is very weak.
Therefore, the important disk parameters for the GI are $f_\mathrm{g}$,
 $f_\mathrm{d}$ and $T_1$. 
We plot $\alpha_\mathrm{cr}$ in Figure \ref{fig:gicond}, which agrees well with the numerical results.

The sufficient condition for the GI is that the dust evolution track
 crosses the GI region.
We can numerically calculate the critical $\alpha$ for the sufficient
 condition.
In our parameter regime $3\, \mathrm{AU} < a < 20\, \mathrm{AU}$ and
 $f_\mathrm{g}>1$, we empirically find that the critical $\alpha$ for
 the sufficient condition is slightly smaller than that for the necessary
 condition as shown in Figure~\ref{fig:gicond}.
The difference is about 50\% at a maximum.
Note that the dust evolution track changes with monomer properties such
 as $r_0$, $\rho_0$, and $E_\mathrm{roll}$.

\section{Summary and Discussion \label{sec:sum}}
We have investigated the stability of the dust disk consisting of
 porous icy dust aggregates using their equilibrium random velocity
 along the compressional evolution due to the self-gravity.
We calculated the equilibrium random velocity considering gravitational
 scattering and collisions among dust aggregates, gas drag, and turbulent
 stirring and scattering.
We obtained the ranges of the mass and internal density of dust
 aggregates for the gravitational instability (GI).
We found that in the minimum-mass solar nebula model with turbulence
 strength $\alpha \lesssim 7\times 10^{-3}$, the disk becomes
 gravitationally unstable as the dust aggregates grow.
The disk with weaker turbulence (smaller $\alpha$) and larger mass
 (larger $f_\mathrm{g}$) is more prone to become gravitationally
 unstable almost independently of its distance from the central star.
For the reasonable ranges of disk parameters the dust evolution
 inevitably leads to the GI.

When the GI occurs, the dust internal density is still low.
The post-GI evolution of the disk of such aggregates was investigated by
 $N$-body simulations \citep{Michikoshi2007, Michikoshi2009, Michikoshi2010}. 
They showed that the GI leads to formation of planetesimals with mass on
 the order of 
\begin{equation}
 m_\mathrm{pl} \simeq
 \lambda_\mathrm{cr}^2 \Sigma_\mathrm{d}
 = 1.42\times 10^{21} f_\mathrm{g}^3
 \left(\frac{f_\mathrm{d}}{0.018}\right)^3
 \left(\frac{a}{5\, \mathrm{AU}}\right) ^{3/2}\, \mathrm{g},
\end{equation}
 where $\lambda_\mathrm{cr} = 4 \pi^2 G \Sigma_\mathrm{d} / \Omega^2$ is
 the critical wavelength of the GI.
We propose the GI of the porous-dust disk as a viable mechanism for planetesimal formation. 
Note that their disk models are rather limited and the further investigation of the post-GI evolution is also necessary.

In the present paper, we adopted the limited disk model and the simple model of the dust aggregate dynamics to see the basic physics as a first step. 
Using more general disk models and more realistic dynamics we systematically investigate the disk stability and obtain more
 rigorous GI conditions in the subsequent paper.

\end{document}